\documentclass[journal,twoside,web]{ieeecolor}
\usepackage{tmi}
\usepackage{graphicx}
\usepackage{subfigure}
\usepackage{bm}
\usepackage{amsmath}
\usepackage{amssymb}
\usepackage{graphicx}
\usepackage{color}
\usepackage{url}
\usepackage{multirow}
\usepackage{booktabs}
\usepackage{multirow}
\usepackage{wrapfig}
\usepackage{caption}
\usepackage{cite}
\usepackage{bbding}
\usepackage{textcomp}
\usepackage[letterpaper=true,colorlinks,bookmarks=false]{hyperref}
\def\BibTeX{{\rm B\kern-.05em{\sc i\kern-.025em b}\kern-.08em
		T\kern-.1667em\lower.7ex\hbox{E}\kern-.125emX}}

\begin{document}

\title{\textbf{ROSE}: A Retinal OCT-Angiography Vessel Segmentation Dataset and New Model}

\author{Yuhui Ma, Huaying Hao, Jianyang Xie, Huazhu Fu, Jiong Zhang, Jianlong Yang, Zhen Wang, Jiang Liu, Yalin Zheng, Yitian Zhao
\thanks{Y.~Ma, H.~Hao, J.~Yang and Y.~Zhao are with Cixi Institute of Biomedical Engineering, Ningbo Institute of Industrial Technology, Chinese Academy of Sciences, Ningbo, China. 
H.~Fu is with Inception Institute of Artificial Intelligence, Abu Dhabi, United Arab Emirates. 
J.~Zhang is with Keck School of Medicine, University of Southern California, Los Angeles, US. 
Z.~Wang is with Department of Neurology, the First Affiliated Hospital of Wenzhou Medical University, Wenzhou, China
J.~Liu is  with Department of Computer Science and Engineering, Southern University of Science and Technology, Shenzhen, China. 
Y.~Zheng is with Department of Eye and Vision Science, University of Liverpool, Liverpool, UK.}
\thanks{ \textit{Y. Ma and H. Hao contributed equally to this work.}}
\thanks{\textit{Corresponding author: Yitian Zhao (yitian.zhao@nimte.ac.cn), and Yalin Zheng (yalin.zheng@liverpool.ac.uk).}}}

\maketitle
\begin{abstract}
Optical Coherence Tomography Angiography (OCTA) is a non-invasive imaging technique that has been increasingly used to image the retinal vasculature at capillary level resolution. However, automated segmentation of retinal vessels in OCTA has been under-studied due to various challenges such as low capillary visibility and high vessel complexity, despite its significance in understanding many vision-related diseases. In addition, there is no publicly available OCTA dataset with manually graded vessels for training and validation of segmentation algorithms. To address these issues, for the first time in the field of retinal image analysis we construct a dedicated Retinal OCTA SEgmentation dataset (ROSE), which consists of 229 OCTA images with vessel annotations at either centerline-level or pixel level. This dataset with the source code has been released for public access to assist researchers in the community in undertaking  research in related topics. 
Secondly, we introduce a novel split-based coarse-to-fine vessel segmentation network for OCTA images (OCTA-Net), with the ability  to detect thick and thin vessels separately. In the OCTA-Net, a split-based coarse segmentation module is first utilized to produce a preliminary confidence map of vessels, and a split-based refined segmentation module is then used to optimize the shape/contour of the retinal microvasculature. We perform a thorough evaluation of the state-of-the-art vessel segmentation models and our OCTA-Net on the constructed ROSE dataset. The experimental results demonstrate that our OCTA-Net yields better vessel segmentation performance in OCTA than both traditional and other deep learning methods. In addition, we provide a fractal dimension analysis on the segmented microvasculature, and the statistical analysis demonstrates significant differences between the healthy control and Alzheimer's Disease group. This consolidates that the analysis of retinal microvasculature may offer a new scheme to study various neurodegenerative diseases. 

\end{abstract}

\section{Introduction}
\begin{figure}[t]
\centering{
\includegraphics[width=0.7\linewidth]{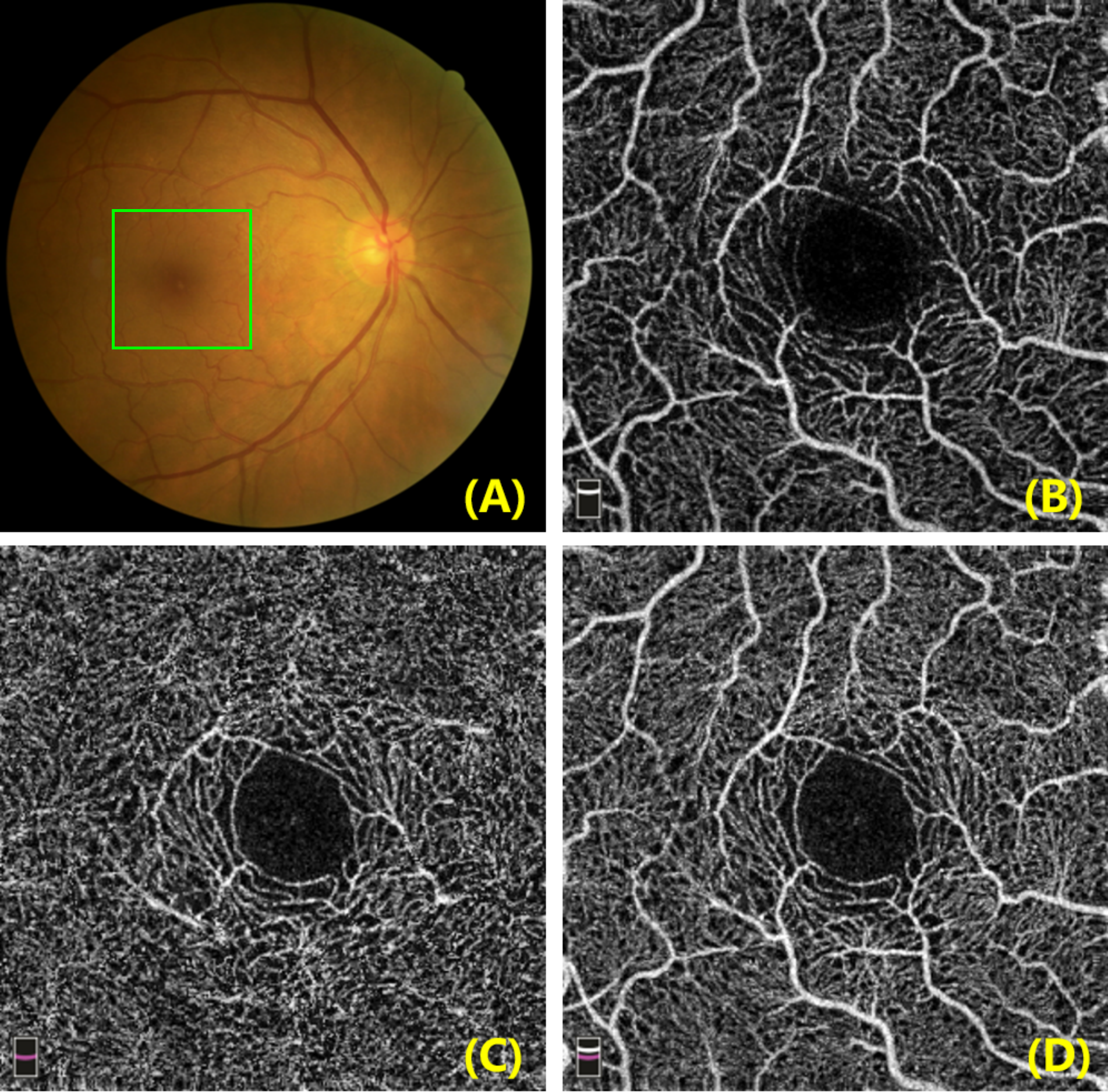}
}
\caption{Illustration of the macula in color fundus image and fovea-centred (green rectangle area) $3\times3~mm^2$ OCTA images of the same eye: (A) color fundus, (B) superficial vascular complexes (SVC), (C) deep vascular complexes (DVC), and (D) the inner retina vascular plexus including both SVC and DVC (SVC+DVC). }
\label{fig.1}
\vspace{-12pt}
\end{figure}

The vasculature is an essential structure in the retina, and its morphological changes can be used not only to identify and classify the severity of systemic, metabolic, and hematologic diseases~\cite{MouMICCAI}, but also to facilitate a better understanding of disease progression, and assessment of therapeutic effects~\cite{ZhaoTMI18}.
Color fundus is the most commonly used retinal imaging technique: however, it is difficult with this method to capture microvasculartures (thin vessels and capillaries), which are surrounded in the fovea and parafovea regions, as shown in the green rectangle area of Fig.~\ref{fig.1} (A). Fluorescein angiography and indocyanine green angiography can resolve the retinal vasculature including capillaries, but they are invasive techniques and may cause severe side effects and even death~\cite{witmer2013comparison}. 

In contrast, Optical Coherence Tomography Angiography (OCTA) is a newly emerging non-invasive imaging technique, with the ability to produce high-resolution 3D images of the retinal vasculature, and has been increasingly accepted as an invaluable imaging tool to observe retinal vessels~\cite{leitgeb2019en,jia2012split-spectrum}. By means of OCTA imaging technology, such as the RTVue XR Avanti SD-OCT system (Optovue, Inc, Fremont, California, USA), equipped with AngioVue software (version 2015.1.0.90), \textit{en face} images of retinal vascular plexus at different depths can be generated by using the maximum projection of OCTA flow signals within specific slabs. 
Fig.~\ref{fig.1} (B-D) show superficial vascular complexes (SVC), deep vascular complexes (DVC), and the inner retinal vascular plexus that includes both SVC and DVC (SVC+DVC). In brief, the SVC extended from 3 $\mu m$ below the internal limiting membrane (ILM) to 15 $\mu m$  below the inner plexiform layer (IPL): the DVC extended from 15 to 70 $\mu m$ below the IPL; and the inner retina extended from 3 $\mu m$ beneath the ILM to the outer boundary of the outer plexiform layer\cite{campbell2017definition}. These plexuses were distinguished and separated automatically, using the proprietary tool supplied with the device.  
OCTA enables observation of microvascular details down to capillary level, permitting quantitative assessment of the microvascular and morphological vessel structures in the retina.

By extracting microvascular structures from different OCTA depth layers, one can obtain their corresponding \textit{en face} projections to analyze their respective variations. In particular, the microvasculature distributed within the parafovea is of great interest, as any abnormality there often indicates the presence of some diseases such as early stage glaucomatous optic neuropathy, diabetic retinopathy, and age-related macula degeneration~\cite{ZhaoTMI2016,eladawi2017automatic,alam2018color,Zhang2019MICCAI,Zhang2020Shape}. 
More recently, several studies have shown that the morphological changes of microvasculature revealed by OCTA are associated with Alzheimer’s Disease and Mild Cognitive Impairment~\cite{delia2017retinal,yoon2019retinal}. A new avenue is thereby opened up to study the relation between the appearance of retinal vessels and various neurodegenerative diseases. 
Thus, automatic vessel detection and quantitative analysis of OCTA images are of great value for the early diagnosis of vascular-related diseases affecting retinal circulation, and the assessment of disease progression. However, automated vessel segmentation in OCTA images has been explored only rarely, and remains a challenging task, despite the fact that many medical segmentation approaches - particularly deep learning based techniques~\cite{fu2016deepvessel,gu2019ce-net} - have achieved great success in segmenting blood vessels.

There is no publicly available OCTA dataset with manual vessel annotations, which hinders the further validation of OCTA segmentation techniques. To our best knowledge, only a few automated methods have been developed to segment the retinal vessels from  OCTA images  based on fully automatic thresholding schemes, such as the methods proposed in~\cite{yousefi2015segmentation,gao2016compensation,camino2018enhanced,sarabi2019an,unknown}. Recently, several deep learning-based methods were developed for vessel segmentation in OCTA images, and each using its own private OCTA dataset. For instances,
 Eladawi \textit{et al.}~\cite{eladawi2017automatic} proposed an automatic method on 47 OCTA images. Zhang \textit{et al.}~\cite{Zhang2020Shape} set up the first 3D OCTA microvascular segmentation approach to directly extract 3D capillary networks from OCTA volume data for subsequent shape modeling and analysis. However, they mainly evaluated the \textit{test re-test reliability} of their 3D framework on 360 OCTA volume images, as there were no manually annotated 3D vessel networks available. Mou \textit{et al.} trained a deep network~\cite{MouMICCAI} with 30 OCTA images, but the small dataset used suggests that this method may not be universally applicable across different pathological scenarios. Li \textit{et al.}~\cite{Li2020OCTA} more recently introduced an image projection network that can achieve 3D-to-2D vessel segmentation: they evaluated their method on 316 OCTA images. 
Although these methods achieve usable segmentations for OCTA vessel analysis, the privacy of their datasets makes a unified evaluation benchmark impossible. 

\begin{figure}[t]
\centering{
\includegraphics[width=0.8\linewidth]{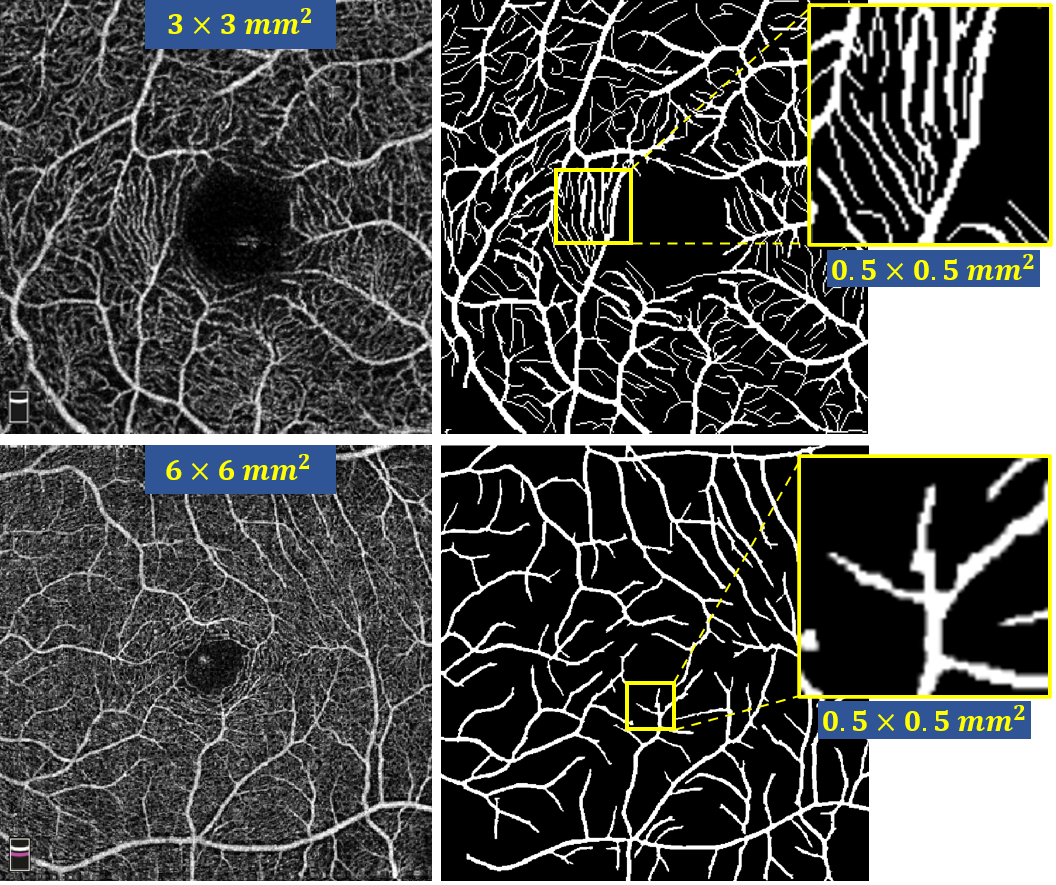}
}
\caption{Comparison of two OCTA images with different scan modes of $3\times 3~mm^2$ (in top row) and $6\times 6~mm^2$ (in bottom row), respectively. Columns 1 to 3 respectively show the original OCTA images, manually annotated vessel networks by experts, and zoomed image patches of the same size ($0.5 \times 0.5~mm^2$) in both scans. }
\label{fig.101}
\vspace{-12pt}
\end{figure}

In addition, Eladawi~\cite{eladawi2017automatic} and Li \textit{et al.}~\cite{Li2020OCTA} set up their segmentation frameworks based on the OCTA data within a $6\times6~mm^2$ fovea-centered field of view (FOV). 
It is worth noting that the scan density for the $3\times3~mm^2$ is higher than that for the $6\times6~mm^2$ scans~\cite{iafe2016retinal}, $6\times6~mm^2$ scans has relatively wider area of scan coverage, and is more like to detect the presence of pathological features, e.g., microaneurysms and non-perfusion. By contrast, the  $3\times3~mm^2$ protocol has a higher scan resolution, and thus is able to delineate the foveal avascular zone (FAZ) and capillaries more clearly than $6\times6~mm^2$ scans.
Previous findings by Zhang~\textit{et al.}~\cite{Zhang2020Shape} have shown that small capillaries play a much more important role in distinguishing different disease groups compared with relatively large vessels. It is therefore necessary to established a dedicated OCTA dataset focusing on more detailed capillary networks within a $3\times3~mm^2$ FOV, and is the main motivation we construct our ROSE dataset. 
Fig.~\ref{fig.101} demonstrates a comparison between $3\times3~mm^2$ and $6\times6~mm^2$ FOVs. We may clearly observe the richer capillary information graded by experts from the $3\times3~mm^2$ scans, while $6\times6~mm^2$ scan has wider FOV but poorer delineation of microvascular. Therefore, the proposed OCTA-Net will be fully evaluated using the well-established $3\times3~mm^2$ OCTA dataset for more detailed microvascular study. 

The OCTA imaging process typically produces images with a low signal-to-noise ratio (SNR)~\cite{jia2012split-spectrum}. Additionally, varying vessel appearances at different depth layers, projection~\cite{hormel2018maximum,zhang2015minimizing},  motion and shadow artifacts~\cite{spaide2015image,chen2016classification,camino2019automated,falavarjani2017image,holmen2020prevalence}, and the potential existence of pathologies~\cite{Zhang2020Shape} increase the challenge for achieving accurate segmentations, particularly for densely connected capillaries, which can easily result in  discontinuous segmentations. Most deep learning-based methods are region-based~\cite{zhang2019et-net}, a technique which is prone to produce imprecise and discontinuous vessel segmentation results, and existing methods do not perform well when required to detect subtle differences in microvascular networks with different vessel thicknesses and imaging depths.

\subsection{Contributions}
In order to mitigate the issues of lacking a public retinal OCTA dataset and effective microvascular segmentation methods, we make the following contributions in this work.


$\bullet$ For the first time in the retinal image analysis field, we establish a publicly available retinal OCTA dataset, with precise manual annotations of retinal microvascular networks, in order to promote relevant research in the community.

$\bullet$ We introduce a novel split-based coarse-to-fine vessel segmentation network for blood vessel segmentation in OCTA, aimed at detecting thick and thin vessels separately. In our method, a split-based coarse segmentation (SCS) module is first utilized to produce a preliminary confidence map of vessels, and a split-based refined segmentation (SRS) module is then used to optimize towards the finer vessels, with a view to obtaining more accurate overall segmentation results.

$\bullet$ We give a full evaluation/benchmarking of OCTA microvascular segmentation, both quantitatively and qualitatively. Comparative analysis shows that the proposed OCTA-Net works robustly on different types of retinal images and yields accurate vessel segmentations.  

To further promote developments in this field, the code, baseline models, and evaluation tools, are publicly available at \textcolor{red}{\url{https://imed.nimte.ac.cn/dataofrose.html}}

\begin{figure*}[t]
\centering 
\includegraphics[width=0.9\linewidth]{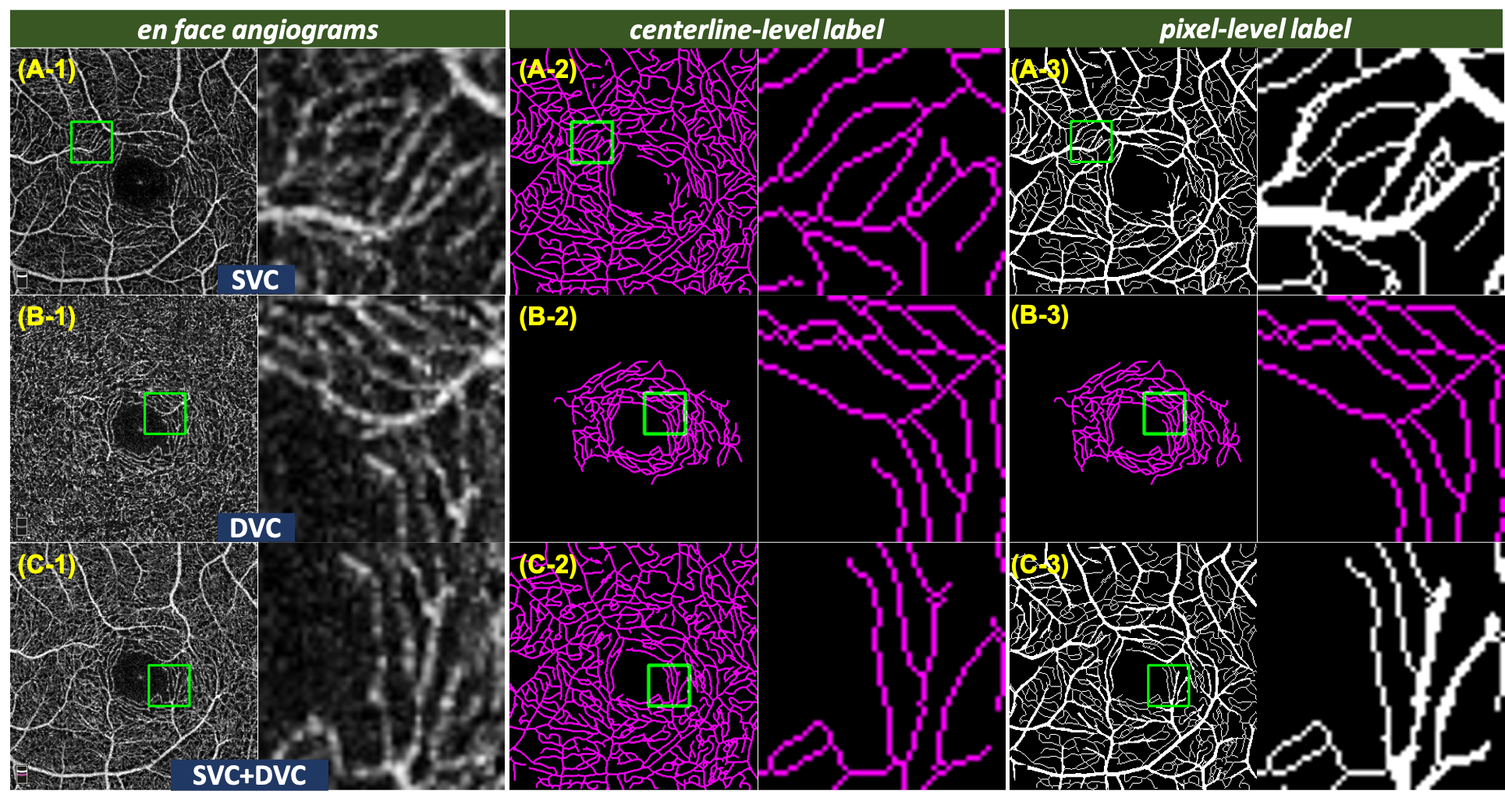} 
\caption{Illustration of OCTA images from \textbf{ROSE-1} and their manual annotations. From top to bottom: \textit{en face} of the SVC, DVC and SVC+DVC, respectively. From left to right: \textit{en face}, centerline-level labels, and pixel-level labels, respectively. }
\label{fig.2}
\vspace{-12pt}
\end{figure*}

\section{Related Works}
In the past two decades, we have witnessed a rapid development of retinal blood vessel segmentation methods for color fundus images (as evidenced by extensive reviews~\cite{Fraz,ZhaoTMI18}). As blood vessels are curvilinear structures distributed across different orientations and scales, the conventional vessel segmentation methods are mainly based on various filters, including Hessian matrix-based filters~\cite{azzopardi2015trainable}, matched filters~\cite{zhang2016robust}, multi-oriented filters~\cite{zhang2017retinal}, symmetry filters~\cite{ZhaoTMI18}, and tensor-based filters~\cite{Cetin2015}. These filter-based methods aim to suppress non-vascular or non-fiber structures and image noise, thus enhance the curvilinear structures, thereby simplifying the subsequent segmentation problem.

By contrast, automated vessel segmentation from OCTA images is relatively unexplored tasks, and most of the existing methods are based on thresholding schemes. Yousefi \textit{et al.}~\cite{yousefi2015segmentation} developed a hybrid Hessian/intensity based method to segment and quantify shape and diameter of vessels. Gao \textit{et al.}~\cite{gao2016compensation} binarized \textit{en face} retinal OCTA images according to mean reflectance projection and maximum decorrelation projection. Camino \textit{et al.}~\cite{camino2018enhanced} set an optimized reflectance-adjusted threshold to segment the vascular. Sarabi \textit{et al.}~\cite{sarabi2019an} developed a three-step algorithm including adaptive thresholding processes to construct the vessel mask. Wu \textit{et al.}~\cite{unknown} proposed an optimized approach based on an improved vascular connectivity analysis (VCA) algorithm to extract the vascular network. However, these thresholding-based approaches are sensitive to noise distributed in \textit{en face} OCTA images and hard to perform well on regions with no significant intensity difference. 

In recent years, deep learning-based methods have made significant progress in the fields of medical image segmentation. In particular, many deep neural networks have been modified and applied for blood vessel segmentation~\cite{Liskowski2016,fu2016deepvessel,alom2018recurrent,MouMICCAI} and have yielded promising results.
However, the extraction of vessels from OCTA images has been relatively unexplored. We will review and discuss the most relevant vessel segmentation works in this section.

A method based on Convolutional Neural Network (CNN)~\cite{Liskowski2016} was proposed to enhance training samples for better retinal vessel detection: subsequently, a Conditional Random Field (CRF) was incorporated into the CNN  by Fu \textit{et al.} for retinal vessel detection~\cite{fu2016deepvessel}. 
Wang \textit{et al.}~\cite{xiancheng2018retina} applied the U-Net~\cite{ronneberger2015u-net:} for retinal vessel segmentation in fundus images of pathological conditions. 
Xiao \textit{et al.}~\cite{xiao2018weighted} modified ResU-Net~\cite{zhang2018road} by introducing a weighted attention mechanism for high-quality retinal vessel segmentation. 
Gu \textit{et al.}~\cite{gu2019ce-net} proposed a context encoder network (CE-Net), which consists of dense atrous convolution and residual multi-kernel pooling modules for retinal vessel image segmentation. 
Jin et.al.~\cite{jin2019dunet:} integrated deformable convolution into the DUNet, which is designed to extract context information and enable precise localization by combining low-level features with high-level ones.
Yan \textit{et al.}~\cite{8476171} proposed a three-stage deep model to segment thick and thin vessels separately, which achieves accurate vessel segmentation for both types of vessels. 
However, there have been very few deep learning methods for vessel segmentation in OCTA images. 
Mou \textit{et al.}~\cite{MouMICCAI} proposed a channel and spatial attention network (CS-Net) for curvilinear structures (including vessels in some example OCTA images) where they applied spatial and channel attention to further integrate local features with their global dependencies adaptively. Li et.al.~\cite{Li2020OCTA} presented an image projection network, which is a novel end-to-end architecture that can achieve 3D-to-2D image segmentation in OCTA images.

\section{Dataset}
Our constructed \textbf{R}etinal \textbf{O}CT-Angiography vessel \textbf{SE}gmentation (\textbf{ROSE}) dataset comprises of two subsets, named as ROSE-1 and ROSE-2, which were acquired by two different devices. All the data described in this paper are acquired from studies that have appropriate approvals from the institutional ethics committees, and written informed consent was obtained from each participant in accordance with the Declaration of Helsinki. The diagnosis result of each subject is provided in the dataset.

\begin{figure}[t]
\centering{
\includegraphics[width=0.8\linewidth]{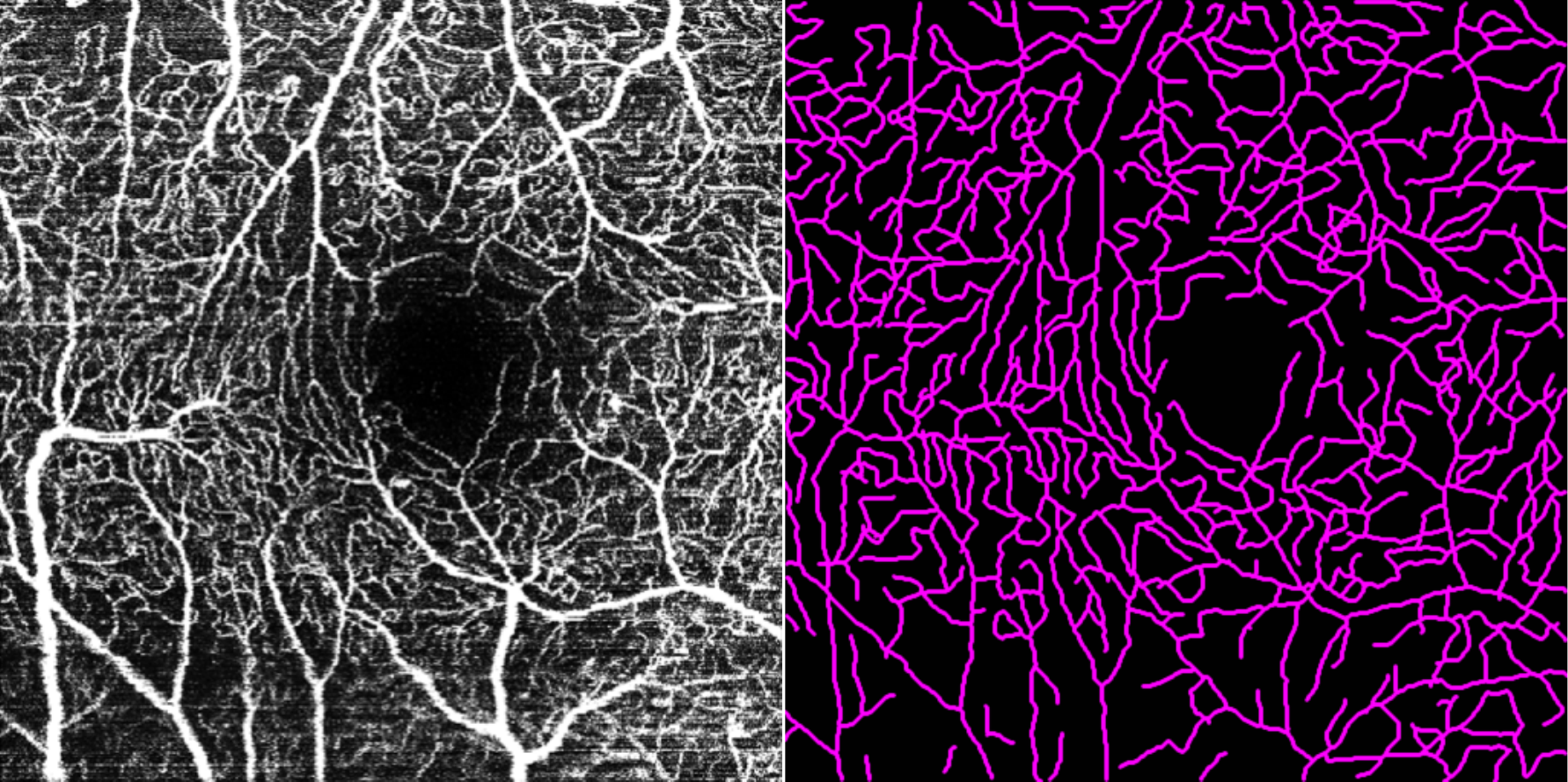}
}
\caption{Example of an OCTA image and its centerline-level manual annotations in \textbf{ROSE-2}.}
\label{roseb_x}
\vspace{-12pt}
\end{figure}

\subsection{ROSE-1}
The ROSE-1 set consists of a total of 117 OCTA images from 39 subjects (including 26 with Alzheimer’s disease (AD) and 13 healthy controls). The mean age was 68.4 $\pm\ 7.4$ years of the AD group and 63.0 $\pm\ 10.2$ years of the control group. Participants with known eye disease, such as glaucoma, age-related macular degeneration, high myopia, etc, and with known systematic disease, e.g., diabetes, were excluded from this study. The diagnosis of AD were based on the  NINCDS-ADRDA criteria~\cite{dubois2007research} and participants did not undergo PET imaging or lumbar puncture for assessment of biomarker status. All the OCTA scans were captured by the RTVue XR Avanti SD-OCT system (Optovue, USA) equipped with AngioVue software, with an image resolution of $304\times304$ pixels. The scan area was $3\times3$ $mm^{2}$ centered on the fovea, within an annular zone of 0.6 $mm$-2.5 $mm$ diameter around the foveal center. The SVC, DVC and SVC+DVC angiograms of each participant were obtained. 

Two different types of vessel annotations were made by image experts and clinicians for the ROSE-1 dataset, and the consensus of them was then used as the ground truth:

\textbf{(1)} \textbf{Centerline-level annotation}. The centerlines of vessels were manually traced using ImageJ software~\cite{schneider2012nih} by our experts on the SVC, DVC, and SVC+DVC images individually, as shown in Fig.~\ref{fig.2} A-2, B-2, and C-2;

\textbf{(2)} \textbf{Pixel-level annotation}. We first invited an image expert to grade the complete microvascular segments with varying diameters in the SVC and SVC+DVC images at pixel level. Since it is difficult for a human expert to perceive the diameters of small capillaries located around the macula region, we asked the expert to grade the small capillaries at centerline level. The combination of these different labels is defined as the final pixel-level annotation, as shown in Fig.~\ref{fig.2} A-3, and C-3. (Note that, Fig.~\ref{fig.2} B-3 is also the centerline-level label of the DVC, as it is difficult to obtain pixel-level grading in this layer.) ROSE-1 dataset were further analysed  in Section Discussion to demonstrate the significance of retinal vasculature in the management of neurodegenerative diseases. 

\begin{figure*}[t]
\centering{
\includegraphics[width=17cm]{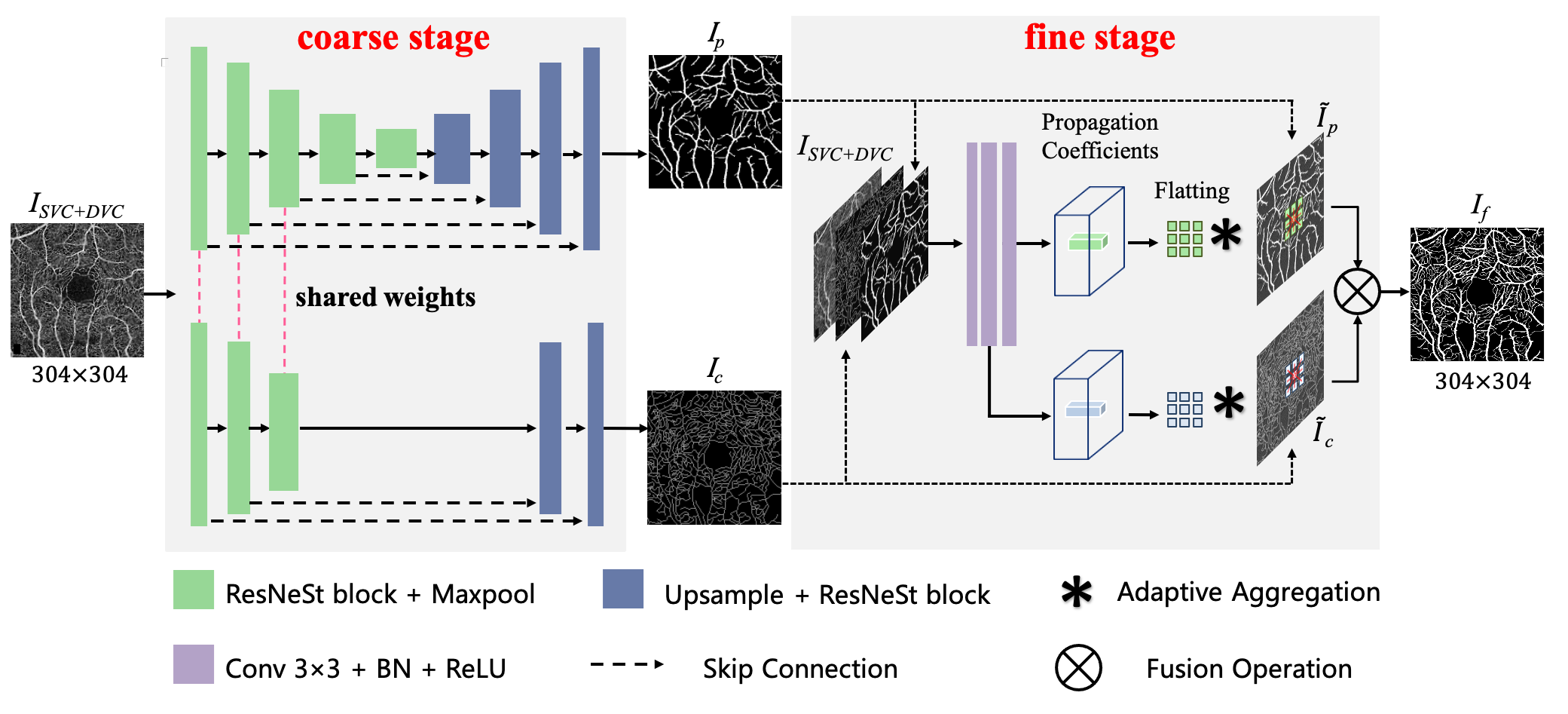}
}
\caption{Architecture of the proposed OCTA-Net network (with an example of \textit{en face} of the angioram of SVC+DVC in the ROSE-1 set). The SCS module (coarse stage) is designed to produce two preliminary confidence maps that segment pixel-level and centerline-level vessels, respectively. The SRS module (fine stage) is then introduced as a fusion network to obtain the final refined segmentation results. }
\label{framework}
\vspace{-12pt}
\end{figure*}
\subsection{ROSE-2}

The ROSE-2 subset contains a total of 112 OCTA images taken from  112 eyes, acquired by a Heidelberg OCT2 system with Spectralis software (Heidelberg Engineering, Heidelberg, Germany). These images are from eyes with various macula diseases. All the images in this dataset are \textit{en face} angiograms of the SVC within a $3\times3$ $mm^{2}$ area centred at the fovea. These OCTA images were reconstructed from $512\times512$ repeated A-scans, with the Heidelberg automated real time (ART) and Trutrack system employed to reduce artefacts and noise. Each image was resized into a grayscale image with $840\times840$ pixels. All the visible retinal vessels were manually traced using an in-house program written in Matlab (Mathworks R2018, Natwick) by an experienced ophthalmologist. An example OCTA image and its corresponding centerline-level annotation are shown in Fig.~\ref{roseb_x}. It should be noted that only the centerlines are annotated at single pixel level for ROSE-2. For better visualization, all the centerlines are widened to 7-pixel wide in the illustration figures. 

\section{Proposed Method}

In this section, we introduce a novel split-based coarse-to-fine network, named as OCTA-Net, for retinal vessel segmentation in OCTA images. The pipeline of OCTA-Net has two indispensable stages - coarse stage and fine stage, as illustrated in Fig.~\ref{framework}. In the coarse stage, a split-based coarse segmentation (SCS) module is designed to produce preliminary confidence maps. In the fine stage, a split-based refined segmentation (SRS) module is used to fuse these vessel confidence maps to produce the final optimized results. 

\subsection{Coarse Stage: Split-based Coarse Segmentation Module}
Since the ROSE-1 set has both pixel-level and centerline-level vessel annotations for each \textit{en face} OCTA image, we design a split-based coarse segmentation (SCS) module with a partial shared encoder and two decoder branches (for pixel-level and centerline-level vessel segmentation, respectively), to balance the importance of both pixel-level and centerline-level vessel information, as illustrated in the coarse stage of Fig.~\ref{framework}. It should be noted that for the ROSE-2 set and \textit{en face} of the DVC layer in the ROSE-1 set on, the designed SCS module only consists of one encoder and one decoder (same architecture as pixel-level vessel segmentation) due to only centerline-level annotations for them.

\begin{figure*}[t]
\centering{
\includegraphics[width=0.8\linewidth]{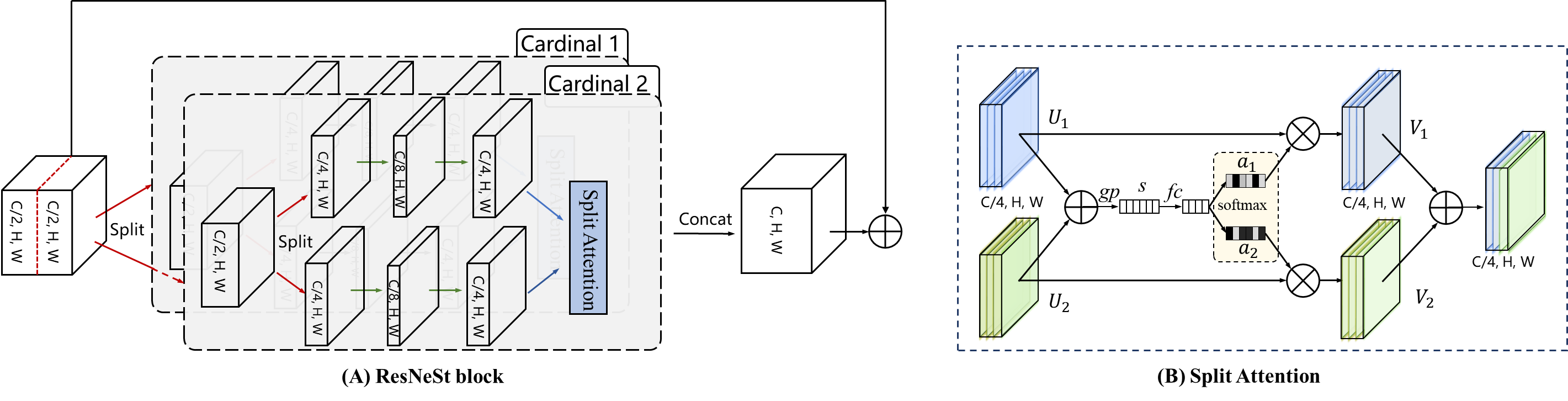}
}
\caption{ResNeSt~\cite{zhang2020resnest} block with split attention module. (A) ResNeSt block. (B) Split attention module. }
\label{resnest}
\vspace{-12pt}
\end{figure*}

\textbf{Pixel-level vessel segmentation: }
Pixel-level vessel segmentation is a U-shape network including five encoder layers and the symmetric decoder layers. A ResNet-style structure with split attention module, ResNeSt block~\cite{zhang2020resnest},  is used as the backbone of each encoder-decoder layer. The primary point of ResNeSt block is to regard a series of representations as the combination of different feature groups, then apply channel-wise attention to these groups. The detailed structure of ResNeSt block is illustrated as Fig.~\ref{resnest} (A). The input $X \in \mathbb{R}^{H \times W \times C}$ of the block is equally split into two cardinal groups which are then fed into two cardinal blocks with same structure respectively. In each cardinal block, the cardinal group is further equally divided and input to two parallel branches. Each branch consists of $1\times1$ and $3\times3$ convolutional layers followed by batch normalization (BN) and ReLU layers, and outputs feature maps with size of $H \times W \times C/4$.

In addition, a split attention module is applied in each cardinal block to integrate these feature maps from the two branches, as illustrated in Fig.~\ref{resnest} (B). The module first fuses feature maps from the two branches (denoted as $U_1$ and $U_2$) via an element-wise summation, then adopts global pooling ($gp$) to generate channel-wise statistics $s$: 

\begin{equation}
\small
s=\frac{1}{H \times W} \sum_{i=1}^{H} \sum_{j=1}^{W} [{U}_{1}(i, j)+{U}_{2}(i, j)]. 
\end{equation}

Two fully connected (FC) layers followed by a softmax layer are then applied to $s$ to obtain $a_{1}$ and $a_{2}$, the channel-wise soft attention weights of $U_1$ and $U_2$ respectively, as illustrated in Fig.~\ref{resnest} (B). 


With the weight vectors $a_{1}$ and $a_{2}$, the weighted results of $U_1$ and $U_2$ and output of the cardinal block $V$ are: $V_{1}=a_{1} \cdot U_{1}$, $V_{2}=a_{2} \cdot U_{2}$, $V=V_{1}+V_{2}$, where $\cdot$ represents channel-wise product. Next, outputs of both cardinal blocks (denoted as $V^1$ and $V^2$) are fused by concatenation along the channel dimension and one $1\times1$ convolution $\mathcal{F}_{1\times1}$: $Z=\mathcal{F}_{1\times1}([V^1, V^2])$. Therefore, the final output Y of the ResNeSt block is produced using a shortcut connection: $Y=Z+T(X)$, where $T$ represents an appropriate transformation, such as stride convolution, combined convolution with pooling or even identity mapping. 

\textbf{Centerline-level vessel segmentation: }
Compared with pixel-level annotation, vessel annotation at centerline level aims to grade the vessels in regions with poor contrast, more complex topological structures, and relatively smaller diameters. On one hand, considering the differences between centerline-level and pixel-level vessels, the features used for pixel-level vessel segmentation might not be suitable for centerline-level vessel segmentation. The deeper architecture might be detrimental to closer attention to low-level features, which are of great significance for centerline-level vessel segmentation. On the other hand, pixel- and centerline-level vessel segmentation may reveal shared features after feature extraction, due to spatial dependencies between the two types of vessel annotations. Based on the above considerations, we append several ResNeSt blocks followed by an up-sampling layer in the third encoder layer of the backbone, as the decoder of the centerline-level vessel segmentation network. Finally, the outputs of the decoder are 
processed by one 1$\times$1 convolutional layer with a Sigmoid function to achieve the centerline-level segmentation map. 

\begin{figure*}[t]
\centering{
\includegraphics[width=0.9\linewidth]{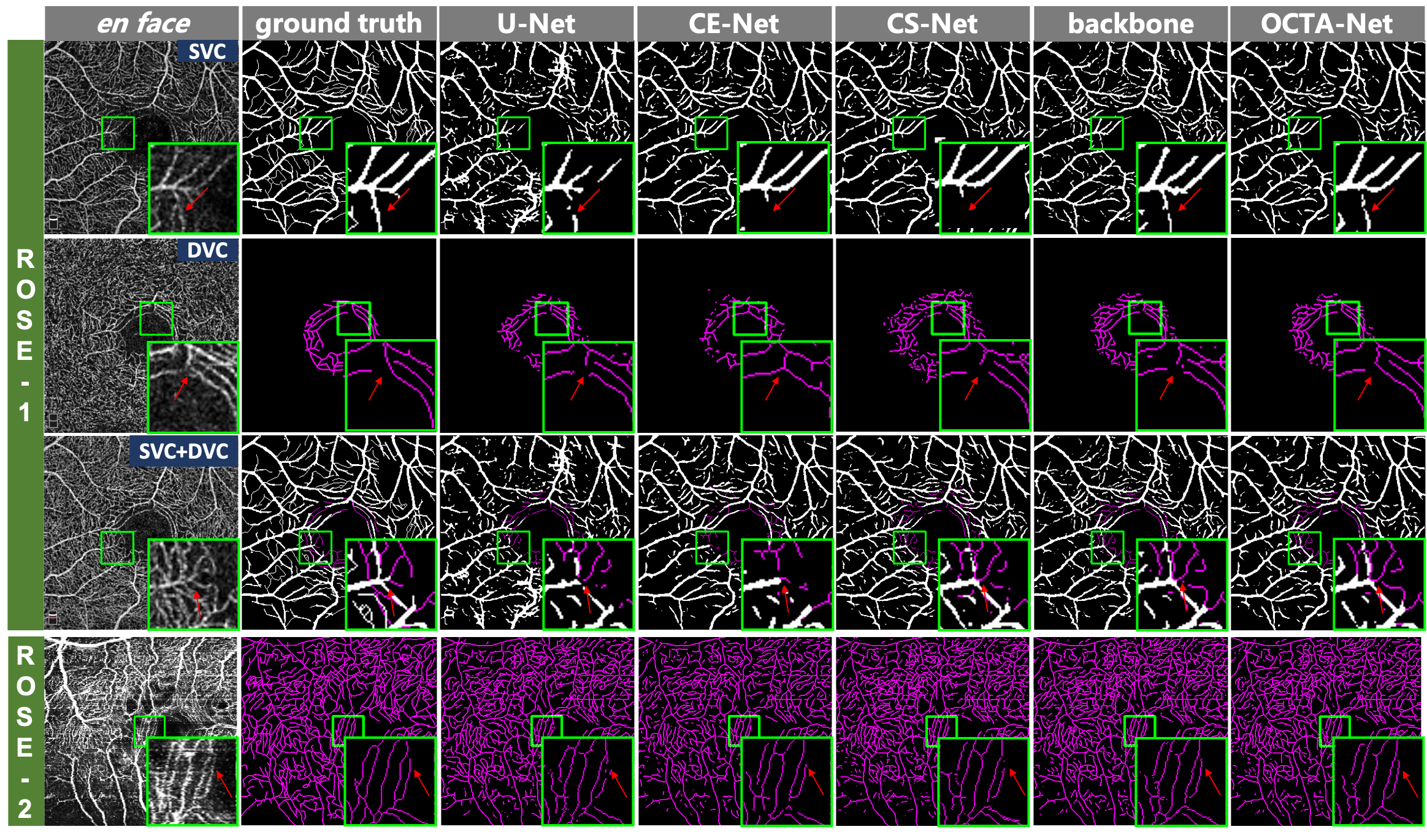}
}
\caption{Vessel segmentation results of different methods on different layers of \textbf{ROSE-1} and \textbf{ROSE-2}. From left to right: \textit{en face} angiograms (original images), manual annotations, vessel segmentation results obtained by U-Net, CE-Net, CS-Net, backbone (ResNeSt) and the proposed method (OCTA-Net), respectively. Note, the color purple indicates the segmentation results at centerline level. }
\label{rosea_results}
\vspace{-12pt}
\end{figure*}

\subsection{Fine Stage: Split-based Refined Segmentation Module}
In order to further recover continuous details of small vessels, we introduce the fine stage to adaptively refine the vessel prediction results{ from the coarse stage}. 
Inspired by~\cite{zhang2017global}, a split-based refined segmentation (SRS) module is proposed as the fine stage. The structure of our SRS module is illustrated in the fine stage of Fig.~\ref{framework}. In order to fully integrate pixel-level and centerline-level vessel information from the SCS module, the predicted pixel-level and centerline-level vessel maps and the original (single channel) OCTA image are first concatenated as input (total 3 channels) to the SRS module. In addition, the SRS module will produce adaptive propagation coefficients to refine the pixel-level and centerline-level maps respectively. In the SRS module, a mini network including three convolutional layers with $3\times3$ kernels is designed to refine the pixel-level map from the coarse stage. Besides, one additional $3\times3$ convolutional layer is appended to the second layer of the mini network to refine the centerline-level map from the coarse stage. BN and ReLU layers are adopted after each convolutional layer. Finally, the refined pixel- and centerline-level maps are then merged into a complete vessel segmentation map, by choosing the larger value from the two maps at each pixel. As in the case of the coarse stage, for the ROSE-2 set and \textit{en face} of the DVC in the ROSE-1 set, no additional convolutional layer is appended to the mini network and only the centerline-level map from the coarse stage is refined. 

The detailed channel configuration of the SRS module is shown in Fig.~\ref{framework}. For vessel refinements, the module produces normalized $m \times m$ local propagation coefficient maps for all the positions, formulated as: 
\begin{equation} 
\small
w_{i}^{p}=\frac{\exp \left(h_{i}^{p}\right)}{\sum_{t=1}^{m \times m} \exp \left(h_{i}^{t}\right)}, \; p \in 1,2, \ldots, m \times m, 
\end{equation}
where $h_{i}^{p}$ is the confidence value at position $i$ for its neighbor $p$, and $m \times m$ is the size of propagation neighbors. Finally, the local propagation coefficient vector $w_{i}^{p}$ at position $i$ will be multiplied by the confidence map of thick or thin vessels from the front model and aggregate to the center point to generate the refinement result, denoted as: 
\begin{equation} 
\small
\setlength{\abovedisplayskip}{2pt}
\setlength{\belowdisplayskip}{2pt}
\mathbf{g}_{i}=\sum_{p=1}^{m \times m} w_{i}^{p} \cdot \mathbf{f}_{i}^{p} ,
\end{equation}
where $\mathbf{f}_{i}^{p}$ is the confidence vector of the neighbor $p$ at position $i$ from the SCS module, and $\mathbf{g}_{i}$ is the final predicted vector at position $i$. 
Note that the propagation coefficient maps can learn the spatial relationship between position $i$ and its neighbors to refine the structure information of vessels. In addition, the final vessel map must be similar to that before refinement. To achieve this goal, the coefficient of position $i$ should be far larger than that of its neighbors, and we adopt a reasonable method for initialization of network parameters following~\cite{zhang2017global}: 
\begin{equation}
\small
\left\{\begin{array}{l}
k_{l}(a, c)=\varepsilon, \varepsilon \sim \mathcal{N}\left(0, \sigma^{2}\right) \text { and } \sigma \ll 1 \\
b_{l}(c)=\left\{\begin{array}{ll}
1 & l=L, c=(m \times m+1) / 2 \\
0 & \text { others }
\end{array}\right. \\
(l=1,2, \ldots, L)
\end{array}\right.
\end{equation}
where $L$ is the number of convolutional layers, $k_l$ and $b_l$ represent convolutional kernels and the bias of layer $l$ respectively, $c$ is the channel of a layer and $a$ is the position in kernels.

\section{Experiments}
\subsection{Experimental Setting}
The proposed method was implemented with PyTorch. Both the coarse and the fine stage were trained with 200 epochs and with the following settings: Adam optimization with the initial learning rate of 0.0005, batch size of 2 and weight decay of 0.0001. For more stable training, we adopted poly learning rate policy with a poly power of 0.9. 

For the coarse stage, we set $r = 16$ as the reduction ratio of the FC layers in the split attention modules, and selected mean square error (MSE) as the loss function: 
\begin{equation} 
\small
\setlength{\abovedisplayskip}{2pt}
\setlength{\belowdisplayskip}{2pt}
L_{MSE} = \frac{1}{N} \sum_{i=1}^{N} \left(p_i - g_i\right)^2 
\end{equation}
where $N$ is the number of all pixels, $p_i$ and $g_i$ represents the $i$-th pixel of the prediction map and the ground truth respectively. 
For the fine stage, we set $m = 3$ as the size of aggregation neighbors. Considering that there is a large imbalance between vessel regions and the background in \textit{en face} OCTA images, we replaced MSE loss with Dice coefficient loss to further optimize the vessel segmentation in the fine stage: 
\begin{equation} 
\small
L_{Dice} = 1 - \frac{2 \sum_{i=1}^{N} p_{i}g_{i} + \epsilon}{\sum_{i=1}^{N} p_{i}^{2} + \sum_{i=1}^{N} g_{i}^{2} + \epsilon} 
\end{equation}
where the parameter $\epsilon$ is a small positive constant used to avoid numerical problems and accelerate the convergence of the training process. 
For training and inference of the proposed method, the ROSE-1 subset was split into 90 images for training and 27 images for testing, and the ROSE-2 subset was split into 90 images for training and 22 images for testing. Data augmentation was conducted by randomly rotation of an angle from $-10^{\circ}$ to $10^{\circ}$ during all training stages. 

We train our OCTA-Net separately for ROSE-1 and ROSE-2, so the annotation differences will not affect the reliability of the evaluation results. In addition, for ROSE-1 dataset, we have both pixel- and centerline-level annotations, so the proposed method learns features from both types of manual annotations.

\subsection{Evaluation Metrics}
To achieve comprehensive and objective assessment of the segmentation performance of the proposed method, the following metrics are calculated and compared:\\ 
$\bullet$ Area Under the ROC Curve (AUC); \\
$\bullet$ Sensitivity (SEN) = TP / (TP + FN); \\
$\bullet$ Specificity (Specificity) = TN / (TN + FP); \\
$\bullet$ Accuracy (ACC) = (TP + TN) / (TP + TN + FP + FN); \\
$\bullet$ $\textit{Kappa}$ score = $(Accuracy - p_e) / (1 - p_e)$; \\
$\bullet$ False Discovery Rate (FDR) = FP / (FP + TP); \\
$\bullet$ $\textit{G-mean}$ score~\cite{ri2020extreme} = $\rm \sqrt{Sensitivity \times Specificity}$; \\  
$\bullet$ Dice coefficient (Dice) = 2 $\times$ TP / (FP + FN + 2 $\times$ TP);\\
where TP is true positive, FP is false positive, TN is true negative, and FN is false negative. $p_e$ in $\textit{Kappa}$ score represents opportunity consistency between the ground truth and prediction, and it is denoted as: 
\begin{equation} 
\small
\begin{aligned} 
\setlength{\abovedisplayskip}{1pt}
\setlength{\belowdisplayskip}{1pt}
p_e = &((TP + FN)(TP + FP) + (TN + FP)(TN + FN)) \\ &/ (TP + TN + FP + FN)^2 
\end{aligned}
\end{equation}

The use of sensitivity and specificity is not adequate for the evaluation of this segmentation task, since over-segmentation still leads to high sensitivity, and the vast majority of pixels do not belong to vessels. Specifically, for centerline-level vessel detection in the DVC images from the ROSE-1 and all images from the ROSE-2, a three-pixel tolerance region around the manually traced centerlines is considered a true positive, which follows the evaluating methods for extracting one pixel-wide curves in~\cite{guimaraes2016a}.

\subsection{Performance Comparison and Analysis} 
We have thoroughly evaluated the proposed method over our ROSE dataset, and compared it to existing state-of-the-art segmentation methods to demonstrate the superiority of our OCTA-Net in the segmentation of OCTA microvasculature.

\begin{table}[t] \centering \scriptsize \caption{Segmentation results obtained using different methods on the {\color{red} SVC} layer from \textbf{ROSE-1}} \setlength{\tabcolsep}{0.7mm}{ 
\begin{tabular}{l||c|c|c|c|c|c|c|c} \hline  \hline   &  \multicolumn{7}{c}{\textbf{ROSE-1 (SVC)}} \\   
\hline 
Methods & AUC & ACC & G-mean & Kappa & Dice & FDR & Time (s) & p-value \\  
\hline 
IPAC~\cite{ZhaoTMI} & 0.8420  & 0.8245 & 0.7517 & 0.4664  & 0.5751 & 0.4816 & - & $<$0.001 \\  
COSFIRE~\cite{azzopardi2015trainable} & 0.9286  & $\textbf{0.9227}$ & 0.7883 & 0.7089  & 0.7517 & $\textbf{0.0471}$ & - & $<$0.01 \\ 
COOF~\cite{Zhang2020Shape} & 0.8689 & 0.8530 & 0.8161 & 0.5684 & 0.6606 & 0.4121 & - & $<$0.001 \\
 U-Net~\cite{ronneberger2015u-net:} &   0.9218&   0.8955 & 0.8068 & 0.6476 &  0.7116&  0.2627 & 0.018 & $<$0.001 \\ 
 ResU-Net~\cite{zhang2018road} &  0.9252&   0.9098 & 0.8229 & 0.6911 &  0.7461&  0.2107 & 0.021 & $<$0.01 \\ 
 CE-Net~\cite{gu2019ce-net} &   0.9292&   0.9121 & 0.8256 & 0.6978 &  0.7511&  0.1995 & 0.025 & $<$0.001 \\
 DUNet~\cite{jin2019dunet:} &   0.9334&  0.9118 & 0.8249 & 0.6970 &  0.7505&  0.2006 & 0.266 & $<$0.001 \\  
 CS-Net~\cite{MouMICCAI} &   0.9392&   0.9152 & 0.8304 & 0.7093 & 0.7608  & 0.1883 & 0.045 & $<$0.001 \\ 
 three-stage~\cite{8476171} &   0.9341&   0.9179 & 0.8318 & 0.7168 &0.7663 & 0.1787 & 33.255 & $<$0.01 \\ \hline 
 OCTA-Net & $\textbf{0.9453}$ & 0.9182 & $\textbf{0.8361}$ & $\textbf{0.7201}$ & $\textbf{0.7697}$ & 0.1775 & 0.059 & - \\  
\hline  
\end{tabular}}   
\label{rosea_SVC}   
\vspace{-10pt}
\end{table}

\begin{table}[t] 
\centering 
\scriptsize
\caption{Segmentation results obtained using different methods on the {\color{red} DVC} images from \textbf{ROSE-1} } \setlength{\tabcolsep}{0.7mm}{ 
\begin{tabular}{l||c|c|c|c|c|c|c|c} \hline  \hline   &  \multicolumn{7}{c}{\textbf{ROSE-1 (DVC)}} \\   
\hline 
Methods & AUC & ACC & G-mean & Kappa & Dice & FDR & Time (s) & p-value \\  
\hline 
IPAC~\cite{ZhaoTMI} & 0.7563  & 0.7522 & 0.7684 & 0.0636  & 0.0911 & 0.9510 & - & $<$0.001 \\  
COSFIRE~\cite{azzopardi2015trainable} & 0.8520  & 0.9130 & 0.8523 & 0.2199  & 0.2405 & 0.8516 & - & $<$0.001 \\ 
 COOF~\cite{Zhang2020Shape} & 0.8162 & 0.6678 & 0.7847 & 0.0649 & 0.1003 & 0.9465 & - & $<$0.001 \\
 U-Net~\cite{ronneberger2015u-net:} & 0.9186  & 0.9895 & 0.8442 & 0.6553 & 0.6605 & 0.3776 & 0.018 & $<$0.01 \\ 
 ResU-Net~\cite{zhang2018road} & 0.9425 & 0.9885 & 0.8553 & 0.6510 & 0.6567 & 0.4002 & 0.021 & $<$0.05 \\ 
 CE-Net~\cite{gu2019ce-net} & 0.9505  & 0.9843 & 0.8503 & 0.5707 & 0.5783 & 0.5147 & 0.025 & $<$0.05 \\
 DUNet~\cite{jin2019dunet:} & 0.9631  & 0.9864 & 0.8369 & 0.6555 & 0.5850 & 0.4930 & 0.266 & 0.105 \\  
 CS-Net~\cite{MouMICCAI} & 0.9671  & 0.9882 & 0.8155 & 0.5825 & 0.5884  & 0.4710 & 0.045 & 0.991 \\ 
  three-stage~\cite{8476171} & 0.9576  & 0.9894 & 0.8399 & 0.6569 & 0.6622  & 0.3780 & 33.255 & $<$0.05\\ \hline 
 OCTA-Net & $\textbf{0.9673}$ & $\textbf{0.9909}$ & $\textbf{0.8811}$ & $\textbf{0.7028}$ & $\textbf{0.7074}$ & $\textbf{0.3492}$ & 0.059 & - \\  
\hline  
\end{tabular}}   
\label{rosea_DVC}   
\vspace{-10pt}
\end{table}

\begin{table}[t] \centering \scriptsize \caption{Segmentation results obtained using different methods on the {\color{red} SVC+DVC} angiograms from \textbf{ROSE-1} } \setlength{\tabcolsep}{0.7mm}{ 
\begin{tabular}{l||c|c|c|c|c|c|c|c} \hline  \hline   &  \multicolumn{7}{c}{\textbf{ROSE-1 (SVC+DVC)}} \\   
\hline 
Methods & AUC & ACC & G-mean & Kappa & Dice & FDR & Time (s) & p-value \\  
\hline 
IPAC~\cite{ZhaoTMI} &  0.7941 &  0.8007 & 0.7054 & 0.3982 &   0.5223&0.5211 & - & $<$0.001 \\  
COSFIRE~\cite{azzopardi2015trainable} &  0.8800 &  0.8981 & 0.7256 & 0.6125 &   0.6671&$\textbf{0.0988}$ & - & $<$0.001 \\ 
COOF~\cite{Zhang2020Shape} & 0.8217 & 0.7762 & 0.7742 & 0.4306 & 0.5685 & 0.5465 & - & $<$0.001 \\
   U-Net~\cite{ronneberger2015u-net:} &  0.9039 &  0.8865 & 0.8050 & 0.6308 & 0.7012 &  0.2892 & 0.018 & $<$0.001 \\ 
   ResU-Net~\cite{zhang2018road} & 0.9108 &  0.8997 & 0.8188 & 0.6689 & 0.7309 &  0.2248 & 0.021 & $<$0.001 \\ 
   CE-Net~\cite{gu2019ce-net} &  0.9155 &  0.8990 & 0.8203 & 0.6678 & 0.7300 & 0.2479 & 0.025 & $<$0.001 \\ 
   DUNet~\cite{jin2019dunet:} &  0.9250 & 0.9046 & 0.8213 & 0.6819 & 0.7403 & 0.2223 & 0.266 & $<$0.001 \\  
   CS-Net~\cite{MouMICCAI} &  0.9311 &  0.9073 & 0.8263 & 0.6919 & 0.7488 & 0.2137 & 0.045 & $<$0.001 \\ 
   three-stage~\cite{8476171} & 0.9248  & 0.9090 & 0.8275 & 0.6967 & 0.7524 & 0.2049 & 33.255 & $<$0.001 \\ 
   \hline 
   OCTA-Net & $\textbf{0.9375}$ & $\textbf{0.9099}$ & $\textbf{0.8338}$ & $\textbf{0.7022}$ & $\textbf{0.7576}$ & 0.2087 & 0.059 & - \\  
\hline  
\end{tabular}}   
\label{rosea_IRVC}
\vspace{-10pt}
\end{table}

\begin{table}[t] \centering \scriptsize \caption{Segmentation results obtained using different methods on \textbf{ROSE-2} } \setlength{\tabcolsep}{0.7mm}{ \begin{tabular}{l||c|c|c|c|c|c|c|c} \hline  \hline  &  \multicolumn{7}{c}{\textbf{ROSE-2}} \\  \hline  Methods & AUC & ACC & G-mean & Kappa & Dice & FDR & Time (s) & p-value \\  \hline  
IPAC~\cite{ZhaoTMI} & 0.7370 & 0.8592 & 0.8207 & 0.4758 & 0.5515 & 0.5590 & - & $<$0.001 \\  
COSFIRE~\cite{azzopardi2015trainable} & 0.7787 & 0.9212 & 0.7742 & 0.5699 & 0.6142 & 0.3891 & - & $<$0.001 \\ 
COOF~\cite{Zhang2020Shape} & 0.7442 & 0.8945 & 0.8117 & 0.5498 & 0.6112 & 0.4620 & - & $<$0.001 \\
U-Net~\cite{ronneberger2015u-net:} & 0.8370 & 0.9319 & 0.8001 & 0.6177 & 0.6564 & 0.3542 & 0.018 & $<$0.001 \\ 
ResU-Net~\cite{zhang2018road} & 0.8413 & 0.9339 & 0.8074 & 0.6348 & 0.6725 & 0.3337 & 0.021 & $<$0.001 \\ 
CE-Net~\cite{gu2019ce-net} & 0.8467 & 0.9377 & 0.8248 & 0.6708 & 0.7066 & \textbf{0.2930} & 0.025 & $<$0.001 \\ 
DUNet~\cite{jin2019dunet:} & 0.8526 & 0.9372 & 0.8213 & 0.6576 & 0.6935 & 0.3132 & 0.266 & $<$0.001 \\  
CS-Net~\cite{MouMICCAI} & 0.8542 & 0.9385 & 0.8235 & 0.6658 & 0.7010 & 0.3025 & 0.045 & $<$0.001 \\ 
three-stage~\cite{8476171} & 0.8590 & 0.9384 & 0.8280 & 0.6671 & 0.7024 & 0.3071 & 33.255 & 0.244 \\ \hline 
OCTA-Net  & \textbf{0.8603} & \textbf{0.9386} & \textbf{0.8315} & \textbf{0.6724} & \textbf{0.7077} & 0.3019 & 0.059 & - \\  \hline  \end{tabular}}   \label{roseb}   \end{table}

\textbf{Comparison methods.} In order to verify the superiority of our method, we compared our method with other  state-of-the-art segmentation methods on both ROSE-1 and ROSE-2, including three conventional methods: infinite perimeter active contour (IPAC)~\cite{ZhaoTMI}, trainable COSFIRE filters~\cite{azzopardi2015trainable}, and curvelet denoising based optimally oriented flux enhancement (COOF)~\cite{Zhang2020Shape} for their effectiveness in detecting vessels with irregular and oscillatory boundaries; and six deep learning approaches: U-Net~\cite{ronneberger2015u-net:}, ResU-Net~\cite{zhang2018road}, CE-Net~\cite{gu2019ce-net}, DUNet~\cite{jin2019dunet:}, CS-Net~\cite{MouMICCAI} and three-stage networks~\cite{8476171}. For~\cite{ZhaoTMI} and~\cite{azzopardi2015trainable}, the parameters were tuned to achieve segmentation results of all vessels in both ROSE-1 and ROSE-2 subsets. For deep learning approaches, all hyper-parameters were manually adjusted to yield the best performances.

\begin{figure*}[t]
\centering{
\includegraphics[width=0.9\linewidth]{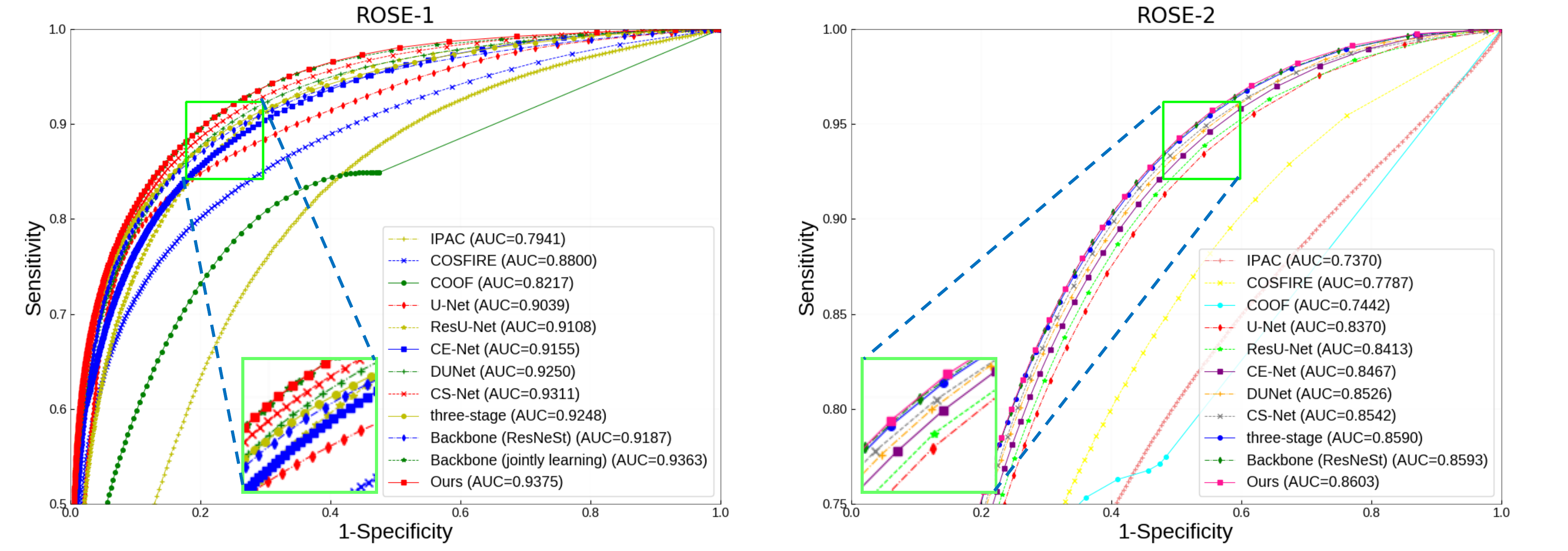}
}
\caption{ROC curves of different methods in terms of SVC+DVC angiograms segmentation performance on ROSE-1 and that of SVC of ROSE-2 datasets. }
\label{rose_roc}
\vspace{-10pt}
\end{figure*}

$\bullet$ \textbf{Subjective comparisons.} Fig.~\ref{rosea_results} presents the respective vessel segmentation results produced by the proposed method, backbone, and the other three selected state-of-the-art segmentation networks.
We can see that U-Net achieves relatively low performance, due to its over-segmentation at regions with high density. CE-Net and CS-Net achieve better performance than U-Net. However, they are not able to preserve fine capillaries well in terms of producing weak vessel responses. In contrast, the proposed method yields more visually informative results. The benefit of the proposed method for segmentation can be observed from the representative regions (green patches). It is clear from visual inspection that our method has identified more complete and thinner vessels particularly in ROSE-1 subset (shown in purple). It achieves relatively uniform responses in both thick and thin vessels, and provides more sensitive and accurate segmentation on capillaries, as demonstrated in the segmentation results of the DVC and SVC+DVC angiograms.

In contrast, all the methods yield very similar segmentation performance in ROSE-2. Therefore, to better evaluate the performance of the proposed method, we provide quantitative results in the following subsections.

$\bullet$ \textbf{Performance on the SVC layer in ROSE-1. }We will first evaluate the vessel segmentation performance on each of the plexus layers of the ROSE-1 subset. Table~\ref{rosea_SVC} quantifies the segmentation performance in SVC images of the different approaches. Overall, our method achieves the best performance in terms of almost all the metrics, with the single exception that its FDR score is 0.0038 lower than that of the three-stage network~\cite{8476171}. Nevertheless, the proposed method is able to correctly identify the majority of vessels using our two-stage architecture. Besides, the last column of Table~\ref{rosea_SVC} also gives paired t-test results in terms of AUC. All $p<0.05$ demonstrate that the proposed method performs significantly better than the other compared methods.

\begin{table*}[t]\footnotesize
 \centering \caption{Ablation studies of our two-stage method on both the ROSE-1 and ROSE-2 subsets} 
 \setlength{\tabcolsep}{1mm}{ 
 \begin{tabular}{l||c|c|c|c|c|c||c|c|c|c|c|c} 
 \hline  
 \hline  
 & \multicolumn{6}{c||}{\textbf{ROSE-1 (SVC+DVC)}} & \multicolumn{6}{c}{\textbf{ROSE-2}}\\  
  \hline  
  Methods &  AUC &ACC & G-mean & Kappa &Dice &FDR&AUC &ACC & G-mean & Kappa & Dice &FDR\\  
  \hline 
 U-Net~\cite{ronneberger2015u-net:} &  0.9039 &  0.8865 & 0.8050 & 0.6308 &  0.7012 &  0.2892 & 0.8370 & 0.9319 & 0.8001 & 0.6177 & 0.6564 & 0.3542 \\ 
  ResU-Net~\cite{zhang2018road} & 0.9108 &  0.8997 & 0.8188 & 0.6689 &  0.7309 &  0.2248 & 0.8413 & 0.9339 & 0.8074 & 0.6348 & 0.6725 & 0.3337 \\ 
  Backbone (ResNeSt) & 0.9187 &  0.9064 & 0.8228 & 0.6887 &  0.7461 & 0.2112 & 0.8593 & 0.9382 & 0.8293 & 0.6692 & 0.7047 & 0.3052 \\ 
 Backbone (joint learning) &  0.9363 &  0.9091 & $\textbf{0.8341}$ & 0.7003 &  0.7564 & 0.2135 & - & - & - & - & - & - \\ 
 \hline 
 Ours & $\textbf{0.9375}$ & $\textbf{0.9099}$ & 0.8338 & $\textbf{0.7022}$ & $\textbf{0.7576}$ & $\textbf{0.2087}$ & \textbf{0.8603} & \textbf{0.9386} & \textbf{0.8315} & \textbf{0.6724} & \textbf{0.7077} & \textbf{0.3019} \\  
\hline 
\end{tabular}}   
\label{ablation}   
\vspace{-10pt}
\end{table*}

$\bullet$ \textbf{Performance on the DVC layer in ROSE-1. }For the DVC images in the ROSE-1 subset, we first adopted the ResNeSt backbone to obtain preliminary vessel segmentation results at the coarse stage. Afterwards, the initial segmentations are combined with their original images as inputs of the fine stage for producing final vessel segmentations. Table~\ref{rosea_DVC} demonstrates the segmentation results achieved by our method and the state-of-the-art methods. Although the improvement on AUC is not significant when compared with CS-Net and DUNet, the proposed network outperformed all the other compared approaches. In particular, it significantly outperforms other methods by a large margin, with an increase of about 12.0\% and 11.9\% in kappa and Dice, respectively, and a reduction of about 13.2\% in FDR when compared with CS-Net. These performance improvements are consistent with the segmentation results shown in the middle row of Fig.~\ref{rosea_results}, where the proposed method successfully extracts small capillaries from macula regions with promising continuity and integrity, while other methods produces relatively lower capillary responses.

$\bullet$ \textbf{Performance on the SVC+DVC angiograms in ROSE-1. }Table~\ref{rosea_IRVC} shows the results of using different segmentation approaches on the SVC+DVC images. Again, the proposed method achieves overall the best performance, with a single exception at the FDR score, where a performance score of 0.0988 is obtained using the method by Azzopardi \textit{et al.}~\cite{azzopardi2015trainable}. However, detection rate of conventional methods of of Azzopardi \textit{et al.}~\cite{azzopardi2015trainable} and Zhao \textit{et al.}~\cite{ZhaoTMI} 
are significantly lower than those of all the deep learning-based methods. This shows that the conventional methods have yet to solve the problems as posed by the high degree of anatomical variations across the population, and the varying scales of vessels within an image. 
Moreover, motion artefacts, noise, poor contrast and low resolution in OCTA exacerbate these problems. By contrast, deep learning-based methods extract a set of higher-level discriminative representations, which are derived from both local and global appearance features and thus can achieve better performance. All $p\leq0.001$ demonstrate the significant performance of our method.

$\bullet$ \textbf{Performance on ROSE-2. }ROSE-2 only provides centerline-level manual annotation, and includes \textit{en face} OCTA images of the SVC. Therefore, as with the DVC images in ROSE-1 subset, for ROSE-2, our method adopts a ResNeSt backbone at the coarse stage to obtain the preliminary vessel segmentation results. Then, the final results are obtained at the fine stage, using the original image and the preliminary results from the coarse stage as input. Table~\ref{roseb} presents the performance of the different segmentation methods. It shows that our method achieves the best AUC, ACC, Kappa and Dice respectively. The statistical analysis reveals that our OCTA-Net outperforms significantly other methods: all $p<0.001$ with a single exception that when compared with three-stage model.

Moreover, Table~\ref{rosea_SVC} -~\ref{roseb} also report the inference time cost of deep learning methods. We don’t list the time cost of three conventional methods here because it is unfair to compare these methods run on CPU or even with MATLAB codes that are not optimized in efficiency with the deep learning methods run on GPU. For fair comparison of inference time, we test all these trained deep learning models with PyTorch. We observed that although consisting of a coarse stage and a fine stage, the proposed two-stage framework is capable to achieve better segmentation performance within a relatively shorter time on both ROSE-1 and ROSE-2.

In order to illustrate the vessel segmentation performance of different methods in a more intuitive manner, we have also provided ROC curves for both the ROSE-1 and ROSE-2 subsets, as shown in Fig.~\ref{rose_roc}. Due to limited space, here we show only the segmentation performances on the SVC+DVC images in ROSE-1. Compared with the conventional methods such as the algorithms proposed by Zhao \textit{et al.}~\cite{ZhaoTMI} and Azzopardi~\textit{et al.}~\cite{azzopardi2015trainable}, deep learning based methods demonstrate their superiority in segmenting OCTA images. This is because the introduction of excellent modules such as ResNet and attention blocks, are usually helpful in improving the AUC score of the encoder-decoder architecture. In addition, there are two reasons that our two-stage architecture achieves the best ROC curve (shown in red in both the subfigures of Fig.~\ref{rose_roc}). Firstly, using ResNeSt as the backbone of the encoder-decoder architecture further strengthens performance for feature extraction, which improves the extraction of vessel information at different complexities. Secondly, the fine stage can adjust local details on the basis of preliminary results from the coarse stage, which additionally refines the segmentation accuracy of the coarse stage.

\section{Discussion and Conclusion}
\subsection{Ablation Studies}
In this paper, the proposed vessel segmentation method consists of a ResNeSt backbone, joint learning from pixel-level and centerline-level vessel segmentation, and two-stage training. To validate the effectiveness of these components, we carried out the following ablation studies. 
U-Net~\cite{ronneberger2015u-net:} is treated as the baseline encoder-decoder method. Then, we gradually evaluated how each of these components affect the results.

$\bullet$ \textbf{Ablation for ResNeSt backbone. }To discuss the performance of the ResNeSt backbone, we compare segmentation performance of the original U-Net, ResU-Net~\cite{zhang2018road} (the modified U-Net with residual blocks in the encoder) and our proposed encoder-decoder architecture (with ResNeSt as the backbone), as shown in Table~\ref{ablation}. For both the ROSE-1 and ROSE-2 subsets, our encoder-decoder architecture with ResNeSt as the backbone achieves the best performance on AUC, ACC, Kappa, Dice and FDR in comparison with the original U-Net and ResU-Net. This indicates that the ResNeSt backbone is superior in feature extraction, which reveals more information about vessels with different characteristics.

$\bullet$ \textbf{Ablation for joint learning at pixel-level and centerline-level vessel segmentation. }In addition, we compared joint learning from pixel-level and centerline-level vessel segmentation with only one segmentation branch (with ResNeSt as the backbone) of all vessels for the SVC+DVC images in the ROSE-1 subset, so as to demonstrate the advantages of joint learning from both pixel-level and centerline-level vessel segmentation. Comparisons of both performance are illustrated in Table~\ref{ablation}. We can observe that joint learning achieves higher scores in terms of AUC, ACC, Kappa and Dice than learning from only one segmentation branch. This suggests that joint learning could help to improve both pixel-level and centerline-level vessel segmentation performance by highlighting the relevant topological distinctions between pixel-level and centerline-level vessels.

$\bullet$ \textbf{Ablation for two-stage training.} Furthermore, we analysed the impact of the fine stage on the coarse stage in our two-stage procedure. At the coarse stage, for the ROSE-1 subset, pixel-level and/or centerline-level vessel segmentation results are treated as the preliminary segmentation results, while for the ROSE-2 subset, vessel segmentation results produced by the ResNeSt backbone are treated as the preliminary  segmentation results. At the fine stage, final vessel segmentation results are derived from both the original images and preliminary results from the coarse stage. Accordingly, we made a comparison between the preliminary  segmentation results of the coarse stage and final vessel segmentation results of the fine stage. As illustrated in Table~\ref{ablation}, final vessel segmentation performance at the fine stage for the most parts shows improvement when compared with results from the coarse stage. The last column of Fig.~\ref{rosea_results} also indicates that some details of microvasculature are optimized in the fine stage.

\subsection{Clinical Evaluation}
It has been suggested that the retina may serve as a window for monitoring and assessing cerebral microcirculation~\cite{lee2019associations} and neurodegeneration conditions~\cite{yoon2019retinal}. OCT images have been utilized to observe neurodegenerative changes occurring in the ganglion cell-inner plexiform layer (GC-IPL) thickness and retinal nerve fiber layer (RNFL) thickness of AD and MCI patients~\cite{de2020optical}. Recently, contributions of vascular biomarkers such as length, density and tortuosity, to the diagnosis of MCI and AD are increasingly recognized~\cite{grewal2018assessment,de2020optical}. OCTA, as an extension of OCT, that can provide in vivo, noninvasive visualization of the retinal vessels in different layers. With the simultaneous occurrence of both neurodegeneration and microvascular changes in the brain, many studies~\cite{london2013the,bulut2018evaluation,lee2019associations} have suggested that the macula microvasculature may provide vital information on the changes in the cerebral microcirculation during the subclinical phase. Changes in the retinal capillary network may indicate the onset and progression of retinopathy, and fractal dimension (FD) is a well-known measure for characterizing the geometric complexity of retinal vasculature that will be a promising biomarker for vascular diseases~\cite{cheung2009quantitative,thomas2014measurement,macgillivray2007fractal} as well as neurodegenerative~\cite{delia2018investigating} and cerebrovascular diseases~\cite{lemmens2020systematic}. In particular, Cheung et al.~\cite{cheung2009quantitative} indicated that the conventional structural measures (e.g., branching angle and vascular tortuosity) represent only one of the many aspects of the retinal vascular geometry and are lack of single global measure that can summarize the branching pattern of the vasculature as a whole. On the other hand, FD reflects the overall complexity~\cite{macgillivray2007fractal}, and has been used as a global measure of the geometric pattern of the retinal vasculature potentially representing the complex branching pattern of the microvasculature~\cite{delia2018investigating}.

\begin{figure}[h]
\centering{
\includegraphics[width=0.9\linewidth]{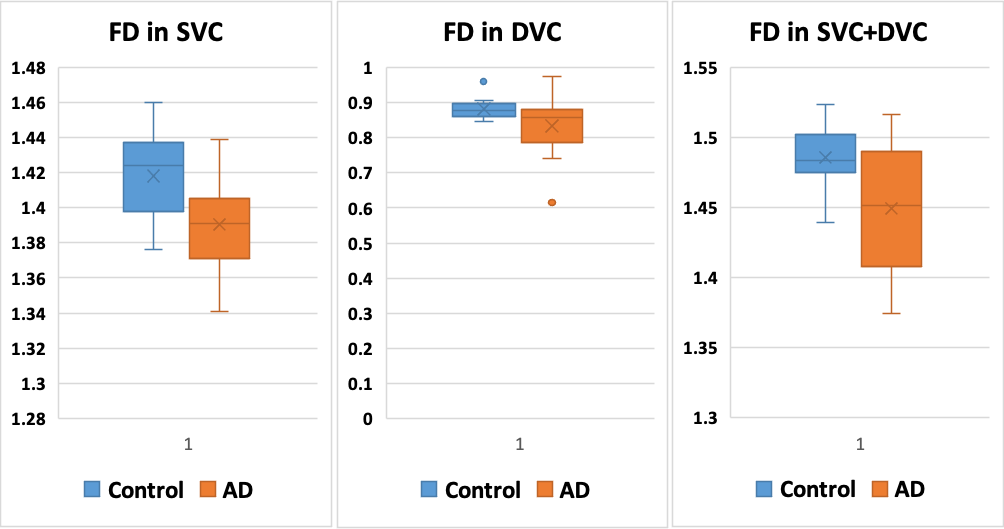}
}
\caption{Descriptive results of the FD of AD patients and controls in the SVC, DVC and SVC+DVC, respectively. }
\label{stat}
\vspace{-10pt}
\end{figure}

In this work, we performed an FD analysis by applying a box-counting method named Fraclab~\cite{falconer1990fractal} on the segmentation results of the SVC, DVC, and SVC+DVC images in ROSE-1 using the proposed method. 
The box-plots in Fig.~\ref{stat} show the statistical analysis results on the ROSE-1 dataset, including 39 images of normal and 78 images of AD subjects, and each subject has three OCTA angiograms: SVC, DVC, and SVC+DVC. We have done three statistical tests with one for each of SVC, DVC and SVC+DVC in order to  avoid the potential correlation issues amongst them. It can be observed that the AD group has reduced FD in the SVC, DVC and SVC+DVC when compared with the control group. Student's t-test was employed to assess the differences between the AD and control groups and results confirmed that the differences between the AD and control participants are significant in the SVC ($p$=0.004$<$0.05), DVC ($p$=0.028$<$0.05) and SVC+DVC ($p$=0.007$<$0.05), respectively. All the three tests have demonstrated significant differences between AD  and healthy groups. These results are consistent with the previous findings that retinal microvascular changes may reflect neurodegenerative changes~\cite{yoon2019retinal,de2020optical}.

\subsection{Projection artefacts}
Projection artefacts are one of the main limitations of OCTA images~\cite{shahlaee2018accentuation}. In this work we used the exported OCTA images as it is and didn't applied any further artifacts removal method. However, Optovue has incorporated their own proprietary three-dimensional projection artifact removal technique into their AngioVue software, and so the ROSE-1 dataset we used in fact have undergone some artifacts removal. In real scenario, projection artefacts may still exist in OCTA images and require further technological improvements, which however is not the focus of the paper. In our experience if a segmentation works on noisy images it will often works well (if not better) in more clean images. We expect that the proposed approach will continue to work on images free of projection artefacts.

\subsection{Conclusions}
In this paper, we have presented a novel Retinal OCTA SEgmentation dataset (ROSE) dataset, which is a large, carefully designed and systematically manually-annotated dataset for vessel segmentation in OCTA images. To our best knowledge, this is the first OCTA dataset released to the research community for the vessel segmentation task. It contains two sub-sets, where the images were acquired by two different devices. All the vessels were manually annotated by human experts at either centerline level and/or pixel level. All of the images contained in the dataset were eventually used for clinical diagnostic purposes. To ensure the utmost protection of patient privacy, the identities of all patients have been removed and cannot be reconstructed. We plan to keep growing the dataset with more challenging situations and various types of eye and neurodegenerative diseases, such as diabetic retinopathy and Parkinson's disease.

In addition to the new dataset, we further proposed a novel two-stage framework for vessel segmentation in OCTA images. In the coarse stage, a split-based coarse segmentation (SCS) module has been designed to achieve the preliminary segmentation results: ResNeSt block is used as the backbone of the framework. In the fine stage, a split-based refined segmentation (SRS) module has been adopted to improve the vessel segmentation results by utilizing both the original images and the preliminary results from the coarse stage. 

The experimental results on the ROSE dataset show that our vessel segmentation approach outperforms other state-of-the-art methods. Small capillaries are major components of the retinal microvasculature and many of them are very tiny segments with only $2-4$ pixel width. Due to the high noise ratio and low capillary visibility in OCTA images, large vessels can be easily extracted while the segmentation of small capillaries becomes challenging. Thus, small increases of metric values may already indicate significant improvement of segmentation with more extracted capillaries. We have conducted paired t-tests on the segmentation performance of our method and other methods. The statistical analysis in Section V.C demonstrates competitive performance of the proposed method. In particular, the improvements of small capillaries in quantitative evaluation metrics are quite significant also due to the extreme imbalance between foreground and background regions.
The sub-analysis on AD shows the great potential of exploring retinal microvascular-based analysis for the diagnosis of various neurodegenerative diseases.

\bibliographystyle{IEEEtran}
\bibliography{refs}

\end{document}